\documentstyle[epsf,epsfig,12pt]{article}
\mark{{}{}}
\def\nll{ \nonumber \\}

\def\lb{\left(}
\def\rb{\right)}

\def\z0{Z}
\def\gf{G_{\mu}}
\def\zm{M_{_Z}}

\def\gev{{\hbox{GeV}}}

\def\wm{M_{_W}}

\def\gz{\Gamma_{_Z}}

\def\barf{\overline f}
\def\barq{\overline q}
\def\barb{\overline b}
\def\bard{\overline d}

\def\barc{\overline c}
\def\bars{\overline s}
\def\baru{\overline u}

\def\barnu{\overline{\nu}}

\def\dr{\Delta r}

\def\i3f{I^{(3)}_f}

\def\osp2{16\,\pi^2}
\def\ap2{\left(p^2\right)}

\def\gev{{\hbox{GeV}}}

\def\s0h{\sigma^h_0}
\def\ba{\begin{eqnarray}}
\def\ea{\end{eqnarray}}

\def\beq{\begin{equation}}
\def\eeq{\end{equation}}
\def\bea{\begin{eqnarray}}
\def\eea{\end{eqnarray}}
\def\barr{\begin{array}}
\def\earr{\end{array}}
\def\bc{\begin{center}}
\def\ec{\end{center}}
\def\btab{\begin{tabular}}
\def\etab{\end{tabular}}

\begin{document}

\begin{flushright}
{\large DFTT 15/96}\\
{\rm April 1996\hspace*{.5 truecm}}\\
\end{flushright}
\vskip 1.cm
\bc{
{\LARGE\bf DISTRIBUTIONS IN FOUR-FERMION\\[.3cm]
 PROCESSES FOR W PHYSICS AT LEP~2}\\[1.cm]
{\large Elena ACCOMANDO, Alessandro BALLESTRERO\\[.2cm]
Giampiero PASSARINO
}}
\vskip 1.cm
{ Dipartimento di Fisica Teorica, Universit\`a di Torino, Italy}\\
{ INFN, Sezione di Torino, Italy}\\
{ v. Giuria 1, 10125 Torino, Italy.}\\
\ec
\noindent
email:
\\
accomando@to.infn.it, ballestrero@to.infn.it, giampiero@to.infn.it

\vskip 3.5cm

\noindent The programs WPHACT and WTO, which are designed 
for computing cross sections 
and other relevant observables in the $e^+e^-$ annihilation into four fermions,
are used to make detailed and complete predictions for the semi-leptonic
and fully hadronic channels $e^+e^- \to \barq q l\nu, \barq q \barq q$.
Both the total cross sections in the LEP~2 energy range and some of
the most relevant distributions are analyzed. Particular algorithms are 
introduced for the fully hadronic channels in order to analyze the $WW$
physics and to properly define the signal versus the background. With
appropriate kinematical cuts it has been shown that the Neutral Current
background can be made vanishingly small when the problem of determining
the $W$ boson mass is addressed. The remaining background from the complete
Charge Current and Mixed processes is again small but not completely 
negligible.
A detailed discussion is performed on the validity of the most relevant
approximations such as the double-resonant one. The inclusion of final
state QCD correction, in its  naive form (NQCD), is discussed and
various implementations are examined.
\newpage

\section{Introduction}

During the last year and in preparation for the experiments to be performed
at LEP~2~\cite{yr}, we have witnessed a new computational
phase in the studies related to the process

\begin{equation}
e^+e^- \to f_1\barf_2 f_3\barf_4.
\end{equation}

\noindent
If one neglects the fermion masses there are $32$ distinct processes of this 
kind (classified in ref.~\cite{wweg}). Many of them have been studied at 
length in the literature, expecially for their interest in the physical
properties of the $W$ boson~\cite{wbos} and of the Higgs boson~\cite{higbos}, 
at energies ranging from that of Lep~1 up to 1~TeV and above.
Several groups have produced Fortran programs 
\cite{wweg}-\cite{fort}-\cite{wto}-\cite{dpeg}, whose results and predictions 
have been  extensively compared during the LEP~2 
Workshop~\cite{wweg}-\cite{dpeg}-\cite{vwg}. 
These codes can be classified  into three broad families, i.e. semi-analytical, 
deterministic and Monte Carlo (MC) integrators, including classical 
event generators (see ref.\cite{wweg}). 
Some of them can produce accurate results for 
(almost) all four fermion processes, with the inclusion of all the relative
Feynman diagrams which
contribute and therefore well beyond some of the most popular and leading 
approximation such as the double resonant one. 

While several different analyses can be performed with one of these programs 
we have concentrated in this paper on the theoretical
predictions which are relevant for the physics of the W boson at Lep~2. 
In any study of this kind, one has to carefully consider both the 
theoretical uncertainties and the interplay between precision measurements
and theoretical calculations.

Some of the published codes have reached an excellent technical agreement 
which is based
on a certain choice of the input parameters, i.e. of the renormalization 
scheme, on the choice of the strategy for 
initial state radiations (structure functions, parton shower, YFS
exponentiation as discussed in appendix A of ref.~\cite{wwcs}), etc. This is 
highly
satisfactory from a technical point of view, but the authors of the codes are
fully aware of the fact that many approximations are still unavoidable at
present. For instance initial and final state radiation are accounted for
only at the level of the leading logarithms (few attempts to go beyond this
approximation can be found in sect. 3.1.14 of ref.~\cite{wweg}), 
QCD corrections are taken into account in a naive way,
which amounts to consider only some of the corrections to the vertices which
become exact for a fully extrapolated setup, no electroweak corrections have 
been included so far, etc. 
This leads to a theoretical uncertainty which at present can only be
roughly estimated by comparing some of the different options. A careful 
analysis on this point has been performed
during the recent LEP~2 Workshop \cite{wweg} and therefore it will not be 
repeated here.

As far as the interplay between precision measurements and theoretical 
calculations is concerned, let us first notice that typically a code will 
produce some differential cross section, including QED corrections and
by using the exact matrix elements for the process, as a function of the 
center-of-mass energy and of the parameters of the vector bosons. It should 
be stressed at this point, that a semi-analytical code is by its own nature 
extremely precise but it allows at most cuts on two out of six of the final 
state invariant masses. However we have at our disposal codes which are not 
only using the exact matrix elements but also which allows for cuts on all 
other variables. 
A crucial role in this context is played by the hadronization 
process~\cite{tb}.
At least in first approximation we could say that the four-fermion codes 
will 
describe the electroweak content of the process to very high accuracy but
they are lacking perturbative parton shower or non-perturbative
hadronization. This raises the question of their reliability for the study
of hadronic or mixed hadronic-leptonic final states. Even though a pragmatic
solution, adopted by many codes,
 consists in standard interfacing with parton-shower and
hadronization programs we still 
insist on the importance of presenting the most precise predictions for cross 
sections and distributions as
they result from the dedicated codes with the inclusion of final states
QCD perturbative corrections. It has been shown by the LEP~1 collaborations
that such predictions are indeed of the upmost importance for understanding
the underlying physical properties of the model once the proper
de-convolution procedure is applied to the data. For this reason we are
still thinking that a correct (theoretical) treatment of the problem at
the level of exact and full matrix elements (including perturbative QCD
corrections) will be essential in understanding several features, not
least the quantitative effect of some of the most common approximations
and the relevance of background versus signal, all of this in presence
of some set of simple but realistic enough cuts.

As already mentioned, in this paper we will analyse the basic properties
of the $W$ boson, 
having in mind expecially but not only the determination of the $W$ boson mass.
It is our opinion that despite the numerous investigations and the various
attempts 
no systematized effort has been spent so far in applying dedicated four-fermion
codes to examine the problem in full detail. We have started from two codes, 
WPHACT 
and WTO~\cite{wto},
which are described at length in the literature and which are based
on completely different methods and approaches. Not only the event generation
is different in the two codes but also the full theoretical framework is 
based on non
intersecting methods. The possibility of comparing results for various
physical observables, all computed with very high numerical accuracy,
gives us an almost absolute confidence on our results and allows us
to give reliable predictions upon which different analyses and strategies
can be based.

Our goal has been therefore to investigate all the Charge-Current 
processes (CC) (typically $\baru d c \bars$)
which are relevant for a measurement of $\wm$ from $WW$ threshold cross
sections and for a direct reconstruction of $\wm$. At the same time
we have been able to perform a detailed study of the corresponding
background induced by Neutral-Current processes (NC) (typically $\baru u
s \bars$)
and by the mixed ones (Mix) (typically $\baru d u \bard$). 
This is a highly non-trivial affair since  the fully hadronic channel
$e^+e^- \to \barq q \barq q$ will receive contributions from $7$ different
types of 
physical processes: one CC ($11$ diagrams),
five NC ($32$ or $64$ diagrams since we include gluons) and one Mix 
($43$ diagrams when again gluons are included). 
We have not included  in our analysis reducible backgrounds such as those due
to four jet processes containing two gluons and two quark jets, which can 
hopefully be distinguished from four quark jets or evaluated with other 
dedicated codes as those used in ref.~\cite{ggcode}.

The outline of the paper is as follows. In Sect. 2 we briefly introduce
the theoretical framework and explain the main conceptual differences
between WPHACT and WTO. In Sect. 3 we concentrate on the semi-leptonic
channel $e^+e^- \to \barq q l\nu$ with particular emphasis on the final
state electron. In Sect. 4 we discuss the fully hadronic channel.
Sect.~5 is devoted to the conclusions.

\section{Theoretical framework}

Ideally the object to be investigated is the cross section for $WW$
production which increases very rapidly near the nominal kinematic
threshold. However we must deal with final states which contain
four fermions in a situation where, for instance, there is no quark-tagging.
Therefore a fully hadronic final state will include all possible
combinations of quarks which is by far more complicated than simply
analyzing $d\baru c\bars$ or similar ones. To understand
the complexity of the problem we start by classifying all the relevant
processes. First the semi-leptonic

\begin{itemize} 

\item $\mu^-\barnu_{\mu}u\bard(c\bars)\,\left[\mu^+\nu_{\mu}\baru d(\barc s)
      \right]$,

\item $e^-\barnu_eu\bard(c\bars)\,\left[e^+\nu_e\baru d(\barc s)\right]$,

\end{itemize}

\noindent
next the fully hadronic ones

\begin{itemize}

\item $u\bard s\barc\,(d\baru c\bars)$,

\item $u\baru d\bard\,(c\barc s\bars)$,

\item $u\baru c\barc$,

\item $u\baru s\bars\,(u\baru b\barb, c\barc d\bard, c\barc b\barb)$,

\item $d\bard s\bars\,(d\bard b\barb, s\bars b\barb)$.

\item $u\baru u\baru\,(c\barc c\barc)$.

\item $d\bard d\bard\,(s\bars s\bars, b\barb b\barb)$.

\end{itemize}

\noindent
The leading contribution below the $ZZ$ threshold is given by those
processes where the fermions can be paired in such a way that they
can derive from a decaying W (even if they do not from the point of
view of Feynman diagrams). Thus for the fully hadronic channel we get
dominant contributions from two processes, each counted with its own
multiplicity,

\begin{equation}
e^+e^- \to d\baru c\bars, \quad d\baru u\bard,
\end{equation}

\noindent
which we term signal while the rest will be referred as background.
Strictly speaking even the signal receive some sort of contamination, since
$d\baru u\bard$ has a part which comes from Neutral-Currents.
Our terminology will be the standard one, therefore 

\begin{itemize}

\item semi-leptonic processes with a $\mu$ are referred as 
 CC10 processes,

\item semi-leptonic processes with a $e$ are referred as CC20 processes,

\item fully hadronic processes are referred as 
 CC11 or NC32 or NC64 or Mix43.

\end{itemize}

\noindent

WPHACT and WTO can deal with all the above channels and for them they can
produce all relevant distributions, from total cross sections to
differential cross sections in energies, scattering angles, invariant
masses or any of their combinations. Both codes can have kinematical
cuts and for that we assumed as a starting point a commonly accepted choice,
the so called canonical cuts which we briefly summarize:

\begin{itemize}

\item $E_l \geq 1\,$GeV, $E_q \geq 3\,$GeV.

\item $M(q_i,q_j), M(q_i,{\bar q}_j), M({\bar q}_i,{\bar q}_j) \geq 5\,$GeV.

\item $10^o \leq \theta_l \leq 170^o$.

\item $\theta(l_i,q_j), \theta(l_i,{\bar q}_j) \geq 5^o$.

\end{itemize}

\noindent
In addition we have selected more restrictive cuts whenever this was of any
relevance for the discussion. 

The long write-up of WPHACT and of WTO~\cite{wto}
can be found in the literature and here we briefly summarize their main 
features.

For WPHACT the code for the full tree level matrix elements for each final
state four fermion process has been written semi-automatically by means of a
set of routines PHACT \cite{Ballestrero} ( Program for Helicity 
Amplitudes Calculations with Tau matrices ) which
implements the helicity formalism of ref.~\cite{BallestreroMaina}. Different
phase spaces are employed to entertain the complex peaking structure of the
Feynman diagrams. 
The adaptive routine VEGAS \cite{vegas} is used for 
integrating over the phase space.
All momenta are explicitly computed in terms of the 
integration
variables and therefore any kinematical cut can be easily performed as well as
distributions for any observable. WPHACT can be also used as a flat event 
generator.
\par
For WTO the helicity amplitudes for each given process
are given, according to the formalism of ref.~\cite{mhf}, in terms of
the $7$ independent invariants which characterize the phase space.
The phase space itself, including all realistic kinematical cuts, is also
described in terms of invariants.  The numerical 
integration, with complete cut-availability, is performed with the help of a 
deterministic integration routine which makes use of quasi-random, 
deterministic number sets, the shifted Korobov sets. The boundaries of the 
phase space, with kinematical cuts, are reconstructed through a backwards 
propagation of constraints.
\par For both codes, initial state QED radiation is included
by means of the structure function approach and upon initialization the final
state QCD corrections are included by adopting a naive approach
(NQCD) to which we will come back later in this section.

There are external blocks present both in WPHACT and in WTO which,
although with a different implementation, have a common root. Among them
we will quote, as most relevant, the choice of the renormalization
scheme and the question of final state QCD corrections.

Four-fermion physics is right now a tree-level prediction and therefore one
can play with the available experimental data points. As an example we observe 
that one of the key relations among the parameters is

\begin{equation}
\gf\wm^2 = {{\pi\alpha}\over {{\sqrt 2}s_{_W}^2}}\,{1\over {1-\dr}}
\end{equation}

\noindent
We have indicated with $\gf$  the Fermi coupling constant, with $\alpha$ 
the fine structure constant and with $s_{_W}$ the sinus of the weak-mixing 
angle. The definitions of the radiative correction factors $\dr$, $\Delta\rho$,
$\Delta r_{rem}$ are given in ref.~\cite{gpyr}.
Given the absence of a full one-loop calculation we have at our disposal 
essentially two non-pathological options, i.e. the so called $\alpha$-scheme

\begin{eqnarray}
s_{_W}^2 &=& {{\pi\alpha}\over {{\sqrt 2}\gf\wm^2}}, \qquad \Delta\rho = 
\Delta r_{rem} = 0,  \\
g^2 &=& {{4\pi\alpha(2\wm)}\over {s_{_W}^2}}
\end{eqnarray}

\noindent
$g$ being the $SU(2)_{_L}$ coupling constant and the so called $\gf$-scheme

\begin{eqnarray}
s_{_W}^2 &=& 1 - {{\wm^2}\over {\zm^2}}, \\
g^2 &=& 4{\sqrt 2}\gf\wm^2, \qquad \Delta\rho = 
\Delta r_{rem} = 0,
\end{eqnarray}

\noindent
which is requested by Ward Identities of the theory. We decided to use the 
latter as our preferred set-up.

To explain our naive treatment of QCD (NQCD) we consider the 
CC10 process $e^+e^- \to \mu^-\barnu_{\mu}u\bard$.
In general one would like to include final state QCD
corrections, even when kinematical cuts are imposed, however a full calculation
is missing. Thus we make use of naive QCD, a simple recipe where the total 
$W$-width is corrected by a factor

\begin{equation}
\Gamma_{_W} \to \Gamma_{_W}\,\lb 1 + \frac{2}{3}\,{{\alpha_s(\wm)}
\over {\pi}}\rb,
\end{equation}

\noindent
where $\alpha_s$ is the strong coupling constant.The cross section 
gets multiplied by a naive factor even in the presence of cuts

\begin{equation}
\sigma_{CC10,C} \to \sigma_{CC10,C}\,\lb 1 + {{\alpha_s(\wm)}
\over {\pi}}\rb.
\end{equation}

\noindent
This naive approach, consequence of our ignorance about the complete
result, would be correct only for $\sigma_{CC03,ex}$, the double-resonant
approximation with fully extrapolated setup. For $\sigma_{CC10,C}$ it
is instead only a rough approximation because of two reasons. First of all in
CC10 we have not only a virtual QCD correction to the $Wu\bard$ vertex but 
also a box diagram. Moreover QED and QCD radiation are quite different if
cuts are imposed, expecially in presence of severe cuts. Thus any inclusion
of final state QCD corrections is, at present, only a very crude approximation
which moreover can become quite bad whenever stringent kinematical cuts are 
applied to the process.

To clearly state the accuracy of our calculations we must add 
that QED radiation
is included by means of the structure function approach (in the so-called
$\beta$-scheme (see appendix A of ref.~\cite{wwcs}). The effect of QED final 
state radiation 
must certainly be included for any reliable determination of the physical 
observables at LEP~2 but we have decided for not including it in the present 
analysis since a more detailed theoretical investigation is needed. 

Before entering the details of our analysis we will give one example of the 
fine-tuning between WPHACT and WTO. Given the cross-section for a specific
choice as $u \bar d s \bar c$
with canonical cuts we have reported in Fig. 1 the relative
deviations from our weighted average as a function of $\sqrt{s}$. The emerging
picture clearly illustrates that we have reached a very high level of
technical agreement. Actually this figure will be the only one where we report
separately results from WPHACT and WTO. For the rest of the paper our results 
should be
interpreted as a common WPHACT/WTO calculation, which have differences
well below what could eventually be appreciated in any realistic figure.
As it will be discussed later there are different implementations of the
naive QCD corrections in
WPHACT and in WTO. Fig. 1a gives the comparison between the preferred 
setups of each code showing differences of order $(\alpha_s/\pi)^2$ whereas 
we have reported in Fig. 1b
the tuned-NQCD comparison clearly showing that below $0.1\%$ everything
really matters.

The exact definition of a vector boson mass, to be extracted from some set of 
data is, to a large extent, dependent on the adopted scheme, i.e. constant
or running width etc. Thus any calculation which aims to determine $\wm$
should clearly state the exact set of conventions under which it has been 
produced. For us a $W$-propagator of invariant mass $s$ is defined by

\begin{equation}
\Delta^{-1}_{_W}(s) = s - \wm^2 + i\,\frac{s}{\wm}\Gamma_{_W}.
\end{equation}

\noindent
corresponding to the so-called running width scheme.

We now specify our set of input-parameters. In the actual calculations we
used

\begin{eqnarray}
\zm &=& 91.1884\,\gev, \qquad \wm= 80.26\,\gev,  \nll
\gz &=& 2.4974\,\gev, \qquad \alpha^{-1}(2\wm) = 128.07.
\end{eqnarray}

\noindent
In both codes the value of $\Gamma_{_W}$ is derived within the minimal standard
model. As for $\alpha(2\wm)$ its value is only relevant for the Coulomb 
correction factor since otherwise we are working in the $\gf$-scheme.
NQCD is implemented according to $\alpha_s(\zm) = 0.123$(input) giving
$\alpha_s(\wm)= 0.1255$.

\section{Semi-leptonic channel $\barq q l\nu$}

This channel, which is relatively easy and clean from a theoretical point
of view, is characterized by the presence of two (or more) hadronic jets, an
isolated energetic lepton and missing energy. We have not taken into account
the $\tau$ as in this case one has a narrow jet due to the hadronic $\tau$
decays. Our cuts will require a threshold energy of $1\,$GeV for the lepton,
with a $10^o$ cuts with respect to the beams. Full angular coverage is
required for the two quarks but their invariant mass has to be greater
than $5\,$GeV. The lepton is also required to be isolated from the hadronic
jets which in our set-up translate into an angular cuts of $5^o$.
We have computed various quantities:

\begin{itemize}

\item the total cross section as a function of $\sqrt s$, from $150\,$GeV
to $205\,$GeV for various choices of the input parameter $\wm$.

\item Several relevant distributions. For the semi-leptonic channel there
is no ambiguity in defining $M_+$ as it can be reconstructed by using
$M(u\bard)$ both in $e^+e^- \to \mu^-\barnu_{\mu}u\bard$ and in
$e^+e^- \to e^-\barnu_eu\bard$. Thus we have computed

\begin{equation}
{{d\sigma}\over {dM_+}},
\end{equation}

\noindent
for a large interval of $M_+$ and for $\sqrt{s} = 161, 175, 190\,$GeV. 
The $\sqrt{s} = 161\,$GeV case has been added for completeness even though
of little experimental interest. 
\end{itemize}

\noindent
The total cross section as a function of $\sqrt{s}$ is reported in
Fig. 2 where we have included a corresponding weight of $4$ which
properly takes into account the following processes

\begin{eqnarray}
e^+e^- &\to& \mu^-\barnu_{\mu} u\bard, \mu^-\barnu_{\mu} c\bars,
\mu^+\nu_{\mu} \baru d, \mu^+\nu_{\mu} \barc s,  \nll
e^+e^- &\to& e^-\barnu_e u\bard, e^-\barnu_e c\bars,
e^+\nu_e \baru d, e^+\nu_e \barc s.
\end{eqnarray}

\noindent
The high energy tail of the figure starts showing a minor difference between
muons and electrons.

The  technical agreement in our predictions enforces the confidence
on $\sigma(E_{cm},\wm)$ upon which one must rely for a $1(2)$-point scan 
needed in the $\wm$ measurement.

In computing the cross section $\sigma(\sqrt{s},\wm)$ we have payed
particular attention to quantities which reflect the sensitivity
to the $W$ mass. In particular we have examined and computed

\begin{equation}
\sigma\mid {{dM}\over {d\sigma}}\mid, \qquad \sqrt{\sigma}\,\mid
{{dM}\over {d\sigma}}\mid, \qquad \mid{{dM}\over {d\sigma}}\mid, 
\end{equation}

\noindent
which contribute to the statistical error and to the systematic errors 
on $\wm$.
All quantities are reported in Fig. 3 as a function of $\sqrt{s} - 2\,\wm$
with a nominal $W$ mass of $80.26$ GeV. In agreement with previous findings
we observe that the statistical sensitivity factor is essentially flat within
$(\sqrt{s})_{min} \pm 2\,$GeV where it varies of approximately $.05\,$GeV 
pb$^{-1/2}$.

For completeness we have shown in Fig. 4 the total cross section for
$e^+e^- \to e^-\barnu_e u \bard$ for different values of $\wm$.

Actually in computing distributions we have been able to compare two rather
different approaches. WPHACT usually collects all the data in a single run
in which a  binning procedure can be automatically started, just
giving the variables for which distributions are  to be evaluated and the
corresponding binning. WTO instead avoids the  binning (even though this
procedure is implemented) and computes directly the differential cross section
by integrating each time over an eight-fold phase space. 
 As a result of the comparison of the two approaches, i.e. something that 
in principle is a  fast
procedure (WPHACT) compared with a  slow but  accurate one (WTO),
we always obtain
that  fast is also  accurate enough in all relevant regions.
Thus we can state that the shape and the moments reconstructed from the 
distributions are
in excellent agreement. From a purely technical point of view, it may be
relevant to notice that the curves of the distributions reported in the 
figures were obtained from WPHACT with a number of bins ranging from $40$ to
more than $100$ per curve. Moreover the statistical errors obtained from
WTO and WPHACT are simply not visible in the plots.

In Fig. 5 we have shown $d\sigma/dM_+$ where
$M_+ = M(u\bard)$ for $\sqrt{s} = 161, 175\,$GeV and $190\,$GeV and for 
$l=e$.  There are no appreciable differences if we consider $l=\mu$, largely
due to our kinematical cuts. We observe that the distribution becomes more and
more symmetric around $M(u \bar d) = \wm$ with growing $E_{cm}$.
From the distributions we have reconstructed four
quantities: the maximum $M_{max}$, the mean $<M>$ and the first moments
$S_{2,3}$. They are reported in the following table

\begin{table}[hbtp]
\begin{center}
\begin{tabular}{|c|c|c|c|c|c|}
\hline
            &                     &     &      &       &     \\
Final state & $\sqrt{s}\,$GeV     & Max  & Mean   & $S_2$ & $S_3$ \\
\hline
            &                     &       &       &      &      \\
$\mu^-\barnu_{\mu}u\bard$ & $161$ & 79.98 & 79.77 & 1.08 & -0.31 \\
            &                     &       &       &      &       \\
$\mu^-\barnu_{\mu}u\bard$ & $175$ & 80.23 & 80.22 & 1.12 & -0.02  \\
            &                     &       &       &      &       \\
$\mu^-\barnu_{\mu}u\bard$ & $190$ & 80.24 & 80.25 & 1.12 & 0.02 \\
            &                     &       &       &      &       \\
$e^-\barnu_eu\bard$ & $161$       & 79.97 & 79.76 & 1.08 & -0.32  \\
            &                     &     &      &       &     \\
$e^-\barnu_eu\bard$ & $175$       & 80.22 & 80.22 & 1.12 & -0.02 \\
            &                     &     &      &       &     \\
$e^-\barnu_eu\bard$ & $190$       & 80.24 & 80.25 & 1.12 & 0.01 \\
            &                     &     &      &       &     \\
\hline
\end{tabular}
\end{center}
\caption{Moments of the $M_+ = M(u\bard)$ distribution.} 
\label{tab1}
\end{table}

In addition, for the semi-leptonic case we have considered the following 
distributions

\begin{equation}
{{d\sigma}\over {dE_{\gamma}}}, \qquad {{d\sigma}\over {dE_l}}, \qquad
{{d\sigma}\over {d\cos\theta_l}},
\end{equation}

\noindent 
which are shown in Fig. 6, 7 and 8 respectively. Within our working scheme

\begin{equation}
E_{\gamma} = \lb 1 - {{x_++x_-}\over 2}\rb\,\sqrt{s}.
\end{equation}

\noindent
where in the c.m.s the $e^{\pm}$ momenta are $P_{\pm} = x_{\pm}p_{\pm}$.

The $E_l$ distribution is again of some relevance in the $\wm$ measurement
since it allows a precise determination of the lepton end-point $E_{\pm}$,

\begin{equation}
E_{\pm} \approx \frac{1}{2}\,E_b \pm \frac{1}{2}\sqrt{E_b^2 - \wm^2},
\end{equation}

\noindent
where $E_b$ is the beam energy.

Few words of comments are in order to explain the relevance of the angular
distributions. For $l=\mu$ the cut imposed on the scattering angle is
irrelevant from a theoretical point of view since we could as well compute
the fully extrapolated cross section 
(CC10 diagrams). For $l=e$ however
it is a completely different story. Here we are dealing with the so called
CC20 diagrams
with $t$-channel photons which induce an apparent singularity at zero
scattering angle. This is of course can be cured by avoiding the approximation
of massless fermions but actually there is more. 

Any calculation for $e^+e^- \to 4$-fermions is only nominally a  tree
level approximation because of the presence of charged and neutral, unstable 
vector bosons and of their interaction with photons.
Unstable particles require
a special care and their propagators, in some channels, must necessarily
include an imaginary part or in other words the corresponding $S$-matrix
elements will show poles shifted into the complex plane. In any 
field-theoretical approach these imaginary parts are obtained by performing
the proper Dyson resommation of the relative two-point functions, which
at certain thresholds will develop the requested imaginary component.
The correct recipe seems representable by a Dyson re-summation of fermionic 
self-energies where only the imaginary parts are actually included. As a 
result the vector boson propagators will be inserted into the corresponding 
tree level amplitudes with a $p^2$-dependent width. It has already been 
noticed by several authors \cite{several}-\cite{gi} that even this simple idea 
gives rise to a series 
of inconsistencies, which sometimes may give results completely inconsistent 
even from a numerical point. The fact is that the introduction of a width into 
the propagators will inevitably result, in some cases, into a breakdown of 
the relevant Ward identities of the theory with a consequent violation of some 
well understood cancellation mechanism. In the CC20 case the effect of 
spoiling a cancellation among diagrams results into a numerical catastrophe
at very small scattering angles.

This simple fact is well illustrated by our calculation where, at various 
energies we have reported in Fig. 9 $d\sigma/d\cos\theta_e$ for $10^o \geq 
\theta_e \geq 1^o$. The two  upper curves (solid and dash) are computed in the
 usual tree level approximation.  Already at $1^o$ we have a growth of two 
order of magnitude with
respect to $10^o$, effect which would become dramatical had we extended our
calculation to smaller angles. Actually we have performed this rather
academic calculation in order to show that WPHACT and WTO agree well even
in some unrealistic and numerically unstable situations. The solution of this
apparent puzzle is by now well know and amounts to adopting the
so-called Fermion-Loop scheme~\cite{gi}. The two lower curves (dotted and 
chain-dot) include  these contributions. 
From the figure can be easily seen however, that 
no appreciable difference between the approximate and correct computation
 is present if a reasonable cut, of about  $5^o$ or greater, is applied.

Moreover we have shown in Fig. 10 the total cross section for the two 
semi-leptonic processes as a function of the cut on the $l^-$ scattering
angle, $\theta_{cut}$. Here one can appreciate the difference between $e$ and
$\mu$ when $\theta_{cut}$ goes to zero.

\section{Fully hadronic channel $\barq q \barq q$}

The fully hadronic channel has a substantial branching ratio and the
typical topology consists of four (or more) energetic jets in the final
state. It has been repeatedly stated in the literature that in the threshold
region the ratio background/signal $\ll 1\%$ but, in this respect,
our analysis represents an  attempt to quantify such a statement.
As already mentioned in the introduction, the background coming from
$2$ gluons and $2$ quarks  and from other processes which
could simulate a $4$ quarks final state have not been considered in our 
computations.

First of all it is important to give a correct definition of  signal
and moreover we need an operative procedure to construct invariant mass
distributions. As for the total cross section we have adopted the following 
algorithm. Let us arbitrarily denote by $i=1,\dots,4$ the four final state 
quarks, then we will have $6$ different invariant masses $M_{ij}, i<j=1,\dots,
4$. We will compute a cross section $\sigma(s)$ by requiring that

\begin{itemize}

\item $M_{12}$ {\tt and} $M_{34}$. {\tt and/or}. $M_{13}$ {\tt and} $M_{24}$.
{\tt and/or}. $M_{14}$ {\tt and} $M_{23}$ are within $10\,$GeV away from
$\wm$ while the remaining invariant masses are above $5\,$GeV.

\end{itemize}

\noindent
In order to define an invariant mass distribution we adopt the following
algorithm

\begin{itemize}

\item For each process we construct 

\begin{equation}
{{d\sigma}\over {dM_i}}, \qquad i=1,\dots,3
\end{equation}

\noindent
where

\begin{enumerate}

\item $M_1 = M_{12} + M_{34}$, $\mid M_{12(34)} - \wm \mid \leq 10\,$GeV

\item $M_2 = M_{13} + M_{24}$, $\mid M_{13(24)} - \wm \mid \leq 10\,$GeV

\item $M_3 = M_{14} + M_{23}$, $\mid M_{14(23)} - \wm \mid \leq 10\,$GeV,

\end{enumerate}

\item those distributions which correspond to 

\begin{enumerate} 

\item $M(d\baru) + M(c\bars)$ in CC11,

\item $M(d\baru) + M(u\bard)$ in Mix43

\end{enumerate}

\noindent
add up, with their multiplicity, to define the signal, while all the
rest is by definition the background.

\end{itemize}

\noindent
In this way we are able to make a quantitative statement on the effect of NC
processes on $WW$ distributions. In the end our procedure amounts to
compute three distributions for $7$ processes, in order to fully account for
the irreducible background to $WW \to \barq q \barq q$.
The main conclusion of our study is that the NC background, $u\baru c\barc$
etc, is completely negligible whenever we apply something of the order
of a $\pm 10\,$GeV cut
around the $W$ mass. The only small but not negligible background is coming
from non-leading contributions of the CC and Mix families. Moreover the leading
contribution of the CC family is completely dominated by the double-resonant
diagrams, the so-called CC03 approximation, at least for the type of cuts that
we have selected. 

Since the effect of the NC processes is marginal whenever a $\pm 10\,$GeV cut 
is applied we can concentrate for a while on the signal, i.e. on the
two processes $e^+e^- \to d\baru c\bars, d\baru u\bard$. From a pure
theoretical point of view in this processes it is possible to identify
$M_{\pm}$ as $M_{c\bars}$ or $M_{u\bard}$. In order to understand the role
of our cuts and the effects of the incorrectly found jet-jet combinations 
we have also compared the total cross sections computed with two different
algorithms. For instance in $e^+e^- \to d\baru u\bard$ we used

\begin{itemize}

\item[A1] $\wm - 10\,$GeV $\leq M(d\baru), M(u\bard) \leq \wm+10\,$GeV,
while $M(d\bard), M(u\baru), M(du)$ and $M(\baru\bard) \geq 5\,$GeV.

\item[A2] $M_{d\baru}$ {\tt and} $M_{u\bard}$. {\tt and/or}. $M_{d\bard}$ 
{\tt and} $M_{u\baru}$. {\tt and/or}. $M_{du}$ {\tt and} $M_{\baru\bard}$ 
are within $10\,$GeV of $\wm$ while the remaining invariant masses are above 
$5\,$GeV.

\end{itemize}

\noindent
Differences are ranging from $1.9\%$ at $\sqrt{s}= 160\,$GeV to $1\%$
at $\sqrt{s} = 175\,$GeV to $1.2\%$ at $\sqrt{s} = 190\,$GeV.
The total cross section with the A2 algorithm is again reported 
in  Fig. 2.
The dash-dotted line refers to the  signal, i.e to the CC11 and Mix43
processes while the solid line gives the total, therefore including all
the NC32+NC64 background which becomes appreciable from energies slightly
below the $ZZ$ threshold.
\par
As far as the multiplicity of all channels is concerned, it is important to
realize  that if one takes into account the mixing induced by the CKM matrix
then several other final states 
come into play, but the net result is just the same as not considering CKM, so
one can really avoid considering these different processes for the present
analysis.
To show this, let us start from the naive case in which there is no CKM mixing
matrix.
Just counting the number of different processes with the same cross sections,
one deduces the following set of weights:

\begin{eqnarray}
W(u\bard s\barc) &=& 2,  \qquad W(u\baru d\bard) = 2,  \nll
W(u\baru c\barc) &=& 1,  \qquad W(u\baru s\bars) = 4,  \nll
W(d\bard s\bars) &=& 3,  \qquad W(u\baru u\baru) = 2,  \nll
W(d\bard d\bard) &=& 3,
\end{eqnarray}

This result is not affected by taking into account CKM mixing matrix.
In fact for CC11, instead of considering only $u\bard s\barc$,
one has now to sum over 

\begin{eqnarray}
u\bard s\barc &(& d\barc, b\barc), \quad 
(V_{cd})^2+(V_{cs})^2+(V_{cb})^2 = 1,  \nll
u\bars s\barc &(& d\barc, b\barc), \quad
(V_{cd})^2+(V_{cs})^2+(V_{cb})^2 = 1,  \nll
u\barb s\barc &(& d\barc, b\barc), \quad
(V_{cd})^2+(V_{cs})^2+(V_{cb})^2 = 1.
\end{eqnarray}

\noindent
Since $(V_{ud})^2+(V_{us})^2+(V_{ub})^2 = 1$ we conclude that the sum of all
these processes gives the same cross section as $u\bard s\barc$.
For mixed processes, instead of the amplitudes for 
$u\bard d\baru$, $u\bars s\baru$, $u\barb b\baru$, one has now to evaluate

\begin{eqnarray}
u\bard d\baru    &{\hbox{Amplitude}}& = NC + V_{ud}^2CC,  \nll
u\bard s\baru    &{\hbox{Amplitude}}& = V_{ud}V_{us}CC,  \nll                  
u\bard b\baru    &{\hbox{Amplitude}}& = V_{ud}V_{ub}CC,  \nll                  
u\bars d\baru    &{\hbox{Amplitude}}& = V_{ud}V_{us}CC,  \nll                  
u\bars s\baru    &{\hbox{Amplitude}}& = NC + V_{us}^2CC,  \nll
u\bars b\baru    &{\hbox{Amplitude}}& = V_{us}V_{ub}CC,  \nll
u\barb d\baru    &{\hbox{Amplitude}}& = V_{ud}V_{ub}CC,  \nll
u\barb s\baru    &{\hbox{Amplitude}}& = V_{us}V_{ub}CC,  \nll
u\barb b\baru    &{\hbox{Amplitude}}& = NC + V_{ub}^2CC,
\end{eqnarray}

\noindent
The sum of the first,second and third $3$ gives for the cross sections

\begin{eqnarray}
&&NC^2+2\,V_{ud}^2NC\times CC+V_{ud}^2\,CC^2,  \nll
&&NC^2+2\,V_{us}^2NC\times CC+V_{us}^2\,CC^2,  \nll
&&NC^2+2\,V_{ub}^2NC\times CC+V_{ub}^2\,CC^2.
\end{eqnarray}

\noindent
The total is $3\,NC^2+2\,NC\times CC+CC^2$ which is as $u\baru d\bard(NC+CC),
u\baru s\bars(NC), u\baru b\barb(NC)$. 
By changing $u \to c$ all processes have been considered and indeed the weights
correspond to the naive ones, without CKM mixing matrix.
\par There is another rather important issue to be discussed, namely to what
extent is the double-resonant approximation (the so called CC03 diagrams) a 
good
approximation. This is entirely cut dependent and by comparing $4\times\,$
CC03 with $2\times\,$(CC11+Mix43) in the A1 algorithm we find very small
differences, of the order of $0.1\%$ from $\sqrt{s} = 160\,$GeV to 
$\sqrt{s} = 205\,$GeV.

The CC03 approximation is an important issue which has been debated
at length. What we claim here is twofold, on one end we have produced
an explicit and complete calculation up to including all fully hadronic
processes (a part from the irreducible background $\bar q q g g$)
where the goodness of the approximation can be  quantitatively
tested. On the other end the goodness of the approximation depends on the
chosen set of cuts and even what can be considered reliable for a study of the
$W$ mass is not also necessarily reliable for the full content of the 
four-fermion physics. 
Moreover the value of the distributions at $M(2j)+M(2j')=M_{max}$   
clearly shows that the CC11 and Mix43  background is
not completely negligible. One last comment concerns the role to be played
by dedicated four-fermion codes. From the LEP~1 experience we know that
one of the possible working options has been to de-convolute the experimental
data and to use the result for fitting the parameters of the standard model.
Of course to create a  4f-fitter requires very high computational speed
associated with reasonably high precision. Both WPHACT and WTO can deal
with semi-leptonic and fully hadronic processes in an efficient way and,
in particular, they are  extremely fast in dealing with CC11-Mix43.
Therefore they both could be interfaced with some fitting procedure
resulting in a fast and accurate determination of the standard model
parameters.

In Fig. 11-13 we have reported the distribution in the  sum of two
invariant masses according to the algorithm previously discussed and
making a distinction between  signal,  CC11 + Mix43 background and
 NC background. The latter has been magnified by a factor of $50$
while the  CC11 + Mix43 one by a factor of $5$. Even if 
this method will probably not be used at $161$ GeV, we report the curves
at such energy in Fig.~11 for completeness.

Our algorithm is based on the general observation that at large $\wm/\sqrt{s}$
one has an excellent determination of $M(2j)+M(2j')$ while $M(2j)-M(2j')$
is poorly determined. Therefore in the $4$-jet channel we never ask which
jet is reconstructed and we use all (three) possible combinations. The
additional cut of $\wm \pm 10\,$GeV simulates in a wide enough mass window
the (almost) equal mass constraint.

This theoretical simulation of the experimental
data handling tells us that the selected cuts are enough to make the
NC background  safely neglected. The remaining effect can
be understood from table~2 where we report some
of the moments for the $M(2j)+M(2j')$ distribution in $e^+e^- \to
2j+2j'$.

\begin{table}[hbtp]
\begin{center}
\begin{tabular}{|c|c|c|c|c|c|}
\hline
            &                     &     &      &       &     \\
Final state & $\sqrt{s}\,$GeV     & Max & Mean & $S_2$ & $S_3$ \\
\hline
            &                     &     &      &       &     \\
total       & $161$               & 159.46 & 157.96 & 2.82 & -2.01 \\
            &                     &        &        &      &     \\
total       & $175$               & 160.40 & 160.34 & 2.95 & -0.44\\
            &                     &        &        &      &     \\
total       & $190$               & 160.42 & 160.54 & 3.03 & 0.25 \\
            &                     &     &      &       &     \\
signal      & $161$               & 159.48 & 158.27 & 2.29 & -1.66\\
            &                     &     &      &       &     \\
signal      & $175$               & 160.40 & 160.36 & 2.48 & -0.26\\
            &                     &     &      &       &     \\
signal      & $190$               & 160.42 & 160.51 & 2.55 & 0.03 \\
            &                     &     &      &       &     \\
\hline
\end{tabular}
\end{center}
\caption{Moments of the $M(2j)+M(2j')$ distribution.} 
\label{tab2}
\end{table}

The previous results obtain for $\wm = 80.26\,$GeV. Within the minimal standard
model and within our renormalization scheme we obtain

\begin{equation}
\Gamma_{_W} = 2.0902\,\gev, \qquad (\alpha_s = 0.1255).
\end{equation}

\noindent
Therefore one easily obtains that in the double-resonant (CC03) approximation
and without initial state QED radiation the maximum of such distributions
should be at

\begin{equation}
M_{max} = 2\,{{\wm^2}\over {\left(\wm^2+\Gamma_{_W}^2\right)^{1/2}}} =
160.47\,\gev.
\end{equation}

\noindent
From table 2 we can easily reconstruct the effect of the background and
of initial state radiation and kinematical cuts.

In this paper we have made no attempt to give a detailed description of
the theoretical uncertainties associated with four fermion production.
However on some specific issue we can point out possible sources of
discrepancy which indeed reflect a theoretical uncertainty. Once we agree
on applying the NQCD prescription we still face two basic options for its
implementation in fully hadronic channels. Given four quarks in the final state
NQCD could amount to multiply by $(1+\alpha_s/\pi)^2$ but one could also
decide to linearize. Since $(\alpha_s/\pi)^2$ is of the order of $1\div 2$
permill the difference will show up (cfr. Fig. 1)
in any comparison which is aimed to
a $0.1\div 0.2$ permill, as the one that we have constantly performed.

Actually there is more in the application of NQCD to fully hadronic
processes. The typical pattern that we have to analyze is the following.
First of all NQCD amounts to neglecting kinematical cuts and to allow
for QCD radiation from external quark lines only. Thus the main approximation
concerns neglecting radiation from internal lines. Even in this approximation
we have quarks which are connected to $W$ and $Z$ bosons, to photons and
to gluons. What to choose for the corresponding scale $\mu$ at which $\alpha_s$
is evaluated? Basically we have made a distinction among three possibilities,
all equally plausible and naive.

\begin{enumerate}

\item For this particular class of processes we fix $\mu$ to be $\wm$ for
CC processes and $\zm$ for NC processes.

\item Still we may choose to apply NQCD everywhere or only in double-resonating
approximation, which means that only $WW$ or $ZZ$ channels are corrected.

\item We adopt a more ambitious program. Each external $\barq q$ pair
is characterized by its invariant mass, no matter where it is coming from,
thus we include NQCD with a correction factor proportional to
$\alpha_s(m_{\barq q})$.

\end{enumerate}

\noindent
For the range of energies and of kinematical cuts implied by the present
analysis it turns out that the three previous options lead to negligibly
small differences.

\section{Conclusions}

Given the intrinsic relevance of having the most reliable predictions
for $W$ physics at LEP~2 energies and strongly motivated by the success
of several numerical comparisons at the recent LEP~2 working groups
we have used two four fermion dedicated FORTRAN codes, WPHACT and WTO, to 
perform a careful a detailed analysis of the distributions in those four 
fermion processes which are relevant at LEP~2 for the measurement of the $W$ 
mass and for the predictions of the standard model concerning the production 
of two $W$ bosons.

The codes use completely different techniques for evaluating the matrix 
elements, for the phase space integration and for producing distributions. 
The perfect (technical) agreement obtained and the smallness of the 
statistical errors enforces the reliability of the results. A word of caution 
should however be spent to recall that these computations are always affected 
by a theoretical uncertainty which has been estimated~\cite{wweg} to be around
a few per mille and which has different roots connected to the choice of
the input parameters ($\alpha_{_{QED}}$ versus $\gf$), of the
renormalization scheme, of the treatment of initial and final state QED 
radiation, of the application of approximate final state QCD correction 
factors. Moreover the computations are always at the parton level: 
no hadronization has been introduced and all cuts refer to the partons.

It is our opinion that, given the uncertainties connected to the hadronization
processes, it is necessary to have the most accurate predictions at the parton
level, in order to disentangle the perturbative regime from the non
perturbative one. When the hadronization programs will be fully tuned also at
LEP~2 and all the problems connected to either colour reconnection or to
Bose-Einstein effect will be completely under control then it will probably 
be possible to de-convolute the data from the hadronization and to compare 
them with dedicated parton level predictions such as those that we have given 
in our analysis and that could eventually form a basis for some fitting
procedure similar the the one which has become so popular at LEP~1. From
this point of view we have shown that all the requirements of computational
speed and of technical precision have been fulfilled.

The results of our study are summarized by the distributions themselves.
They confirm and complete some of the results reported in ref.~\cite{wwmas} by
using a different and complementary approach. 
Moreover, we have proposed some particular algorithm for the fully hadronic 
channels in order to give an unambiguous definition of the signal, of the 
irreducible four-quarks background and of the procedure to construct 
invariant mass distributions even in the absence of flavor reconstruction.

We have shown that with appropriate kinematical cuts one can actually dispose 
of the Neutral Current background when the problem of determining the $W$ 
boson mass is addressed. On the other hand, the background from the complete 
Charge Current and Mixed processes is not completely negligible. We have also 
considered all processes induced by the CKM mixing and concluded that 
introducing them is superfluous for this kind of analyses.

For processes which have an electron in the final state, the so called CC20
diagrams, we have carefully addressed the questions related to numerical
instability and to gauge restoration, therefore giving one of the few
practical and reliable implementations for these processes at small
scattering angle of the electron also confirming that an angular cut around
$10^o$ down to approximately $5^o$ will suffice in guaranteeing
reliable predictions even without having to use a gauge restoring scheme.


\newpage
\centerline{
\epsfig{figure=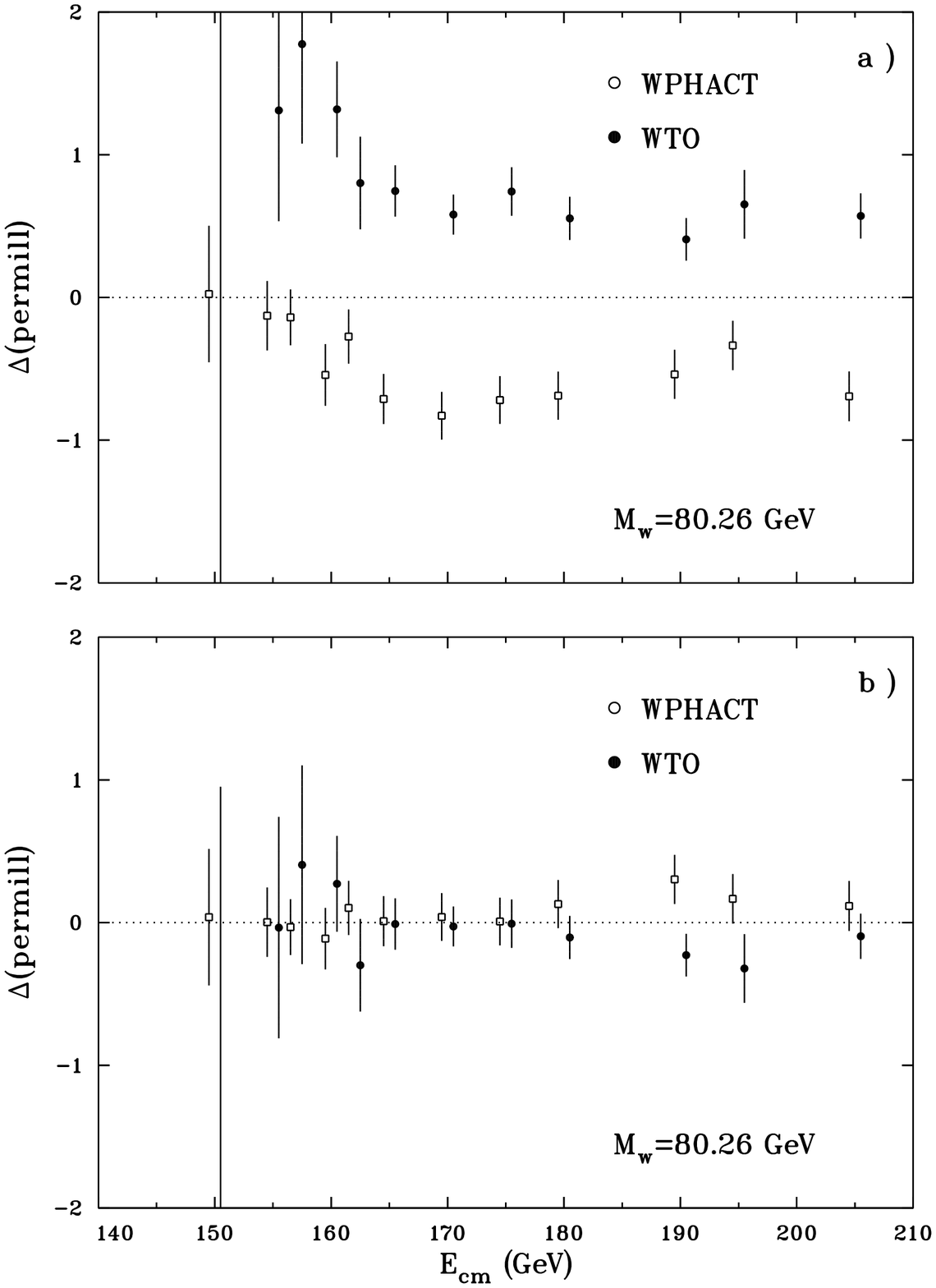,height=20cm,angle=0}
}
Fig.1-  Deviations of the WPHACT and WTO results, 
 from their weighted
average versus $\sqrt{s\,}$, for $u\bar{d}s\bar{c}$ cross-section. 
The upper plot corresponds to the comparison with
NQCD implementations differing of order $(\alpha_s/\pi)^2$, the lower
to the same NQCD implementations.

\newpage
\centerline{
\epsfig{figure=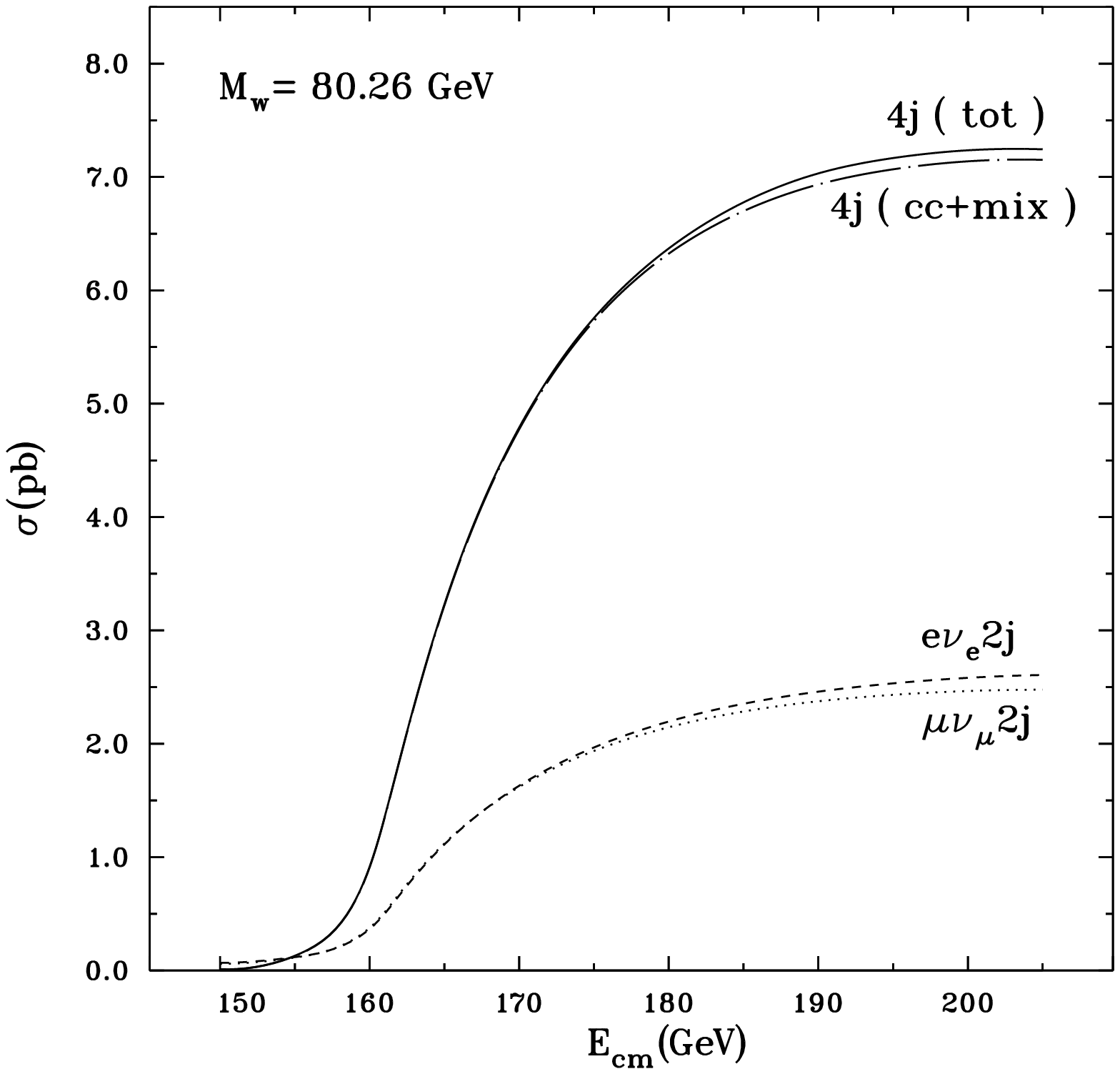,height=18cm,angle=0}
}
Fig.2-  Total cross-section versus $\sqrt{s\,}$, 
for the semi-leptonic channels $q\bar{q}e\nu_e$ (dashed line) and 
$q\bar{q}\mu\nu_{\mu}$
(dotted line) with canonical cuts, and for the fully hadronic
$q_1\bar{q}_2q_3\bar{q}_4$ {\it{signal}} (chaindot line) and
{\it{signal}}+NC {\it{background}} (solid line) with the constraints: a) 
$E_i>3$ GeV, i=1..4, ~b) $M_{12}$ and $M_{34}$, 
and/or, $M_{13}$ and $M_{24}$, and/or, 
$M_{14}$ and $M_{23}$ within 10 GeV away from $\wm$, ~c) 
the remaining invariant masses above 5 GeV.

\newpage
\centerline{
\epsfig{figure=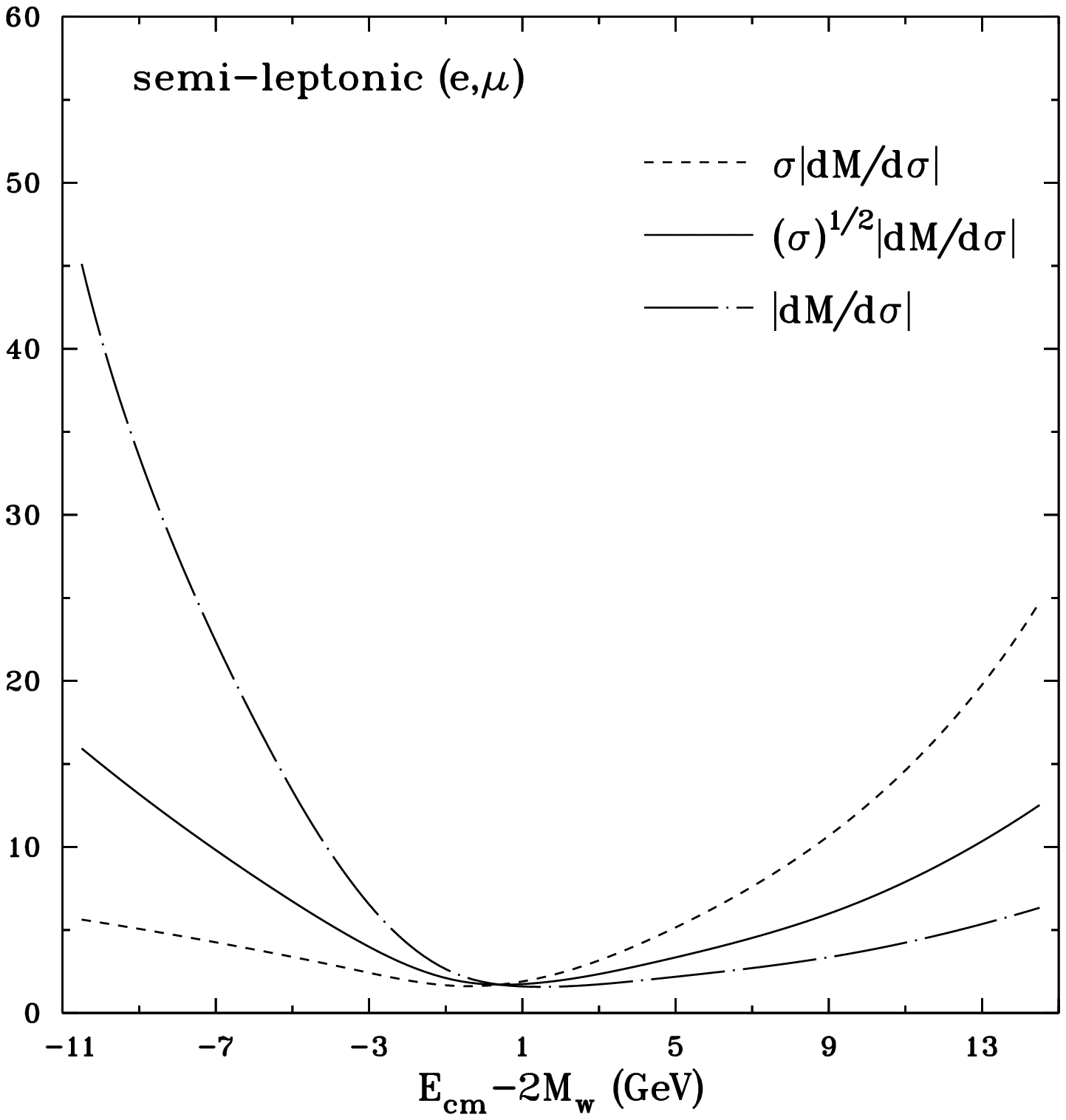,height=18cm,angle=0}
}
Fig.3-  Statistical and systematic sensitivity factors to the W mass 
in the semi-leptonic channel as a function of 
$\sqrt{s\,}-2\wm$, for $\wm = 80.26\,$ GeV. The connection of the three curves
 with the threshold measurement of $\wm$ is discussed in the text. 

\newpage
\centerline{
\epsfig{figure=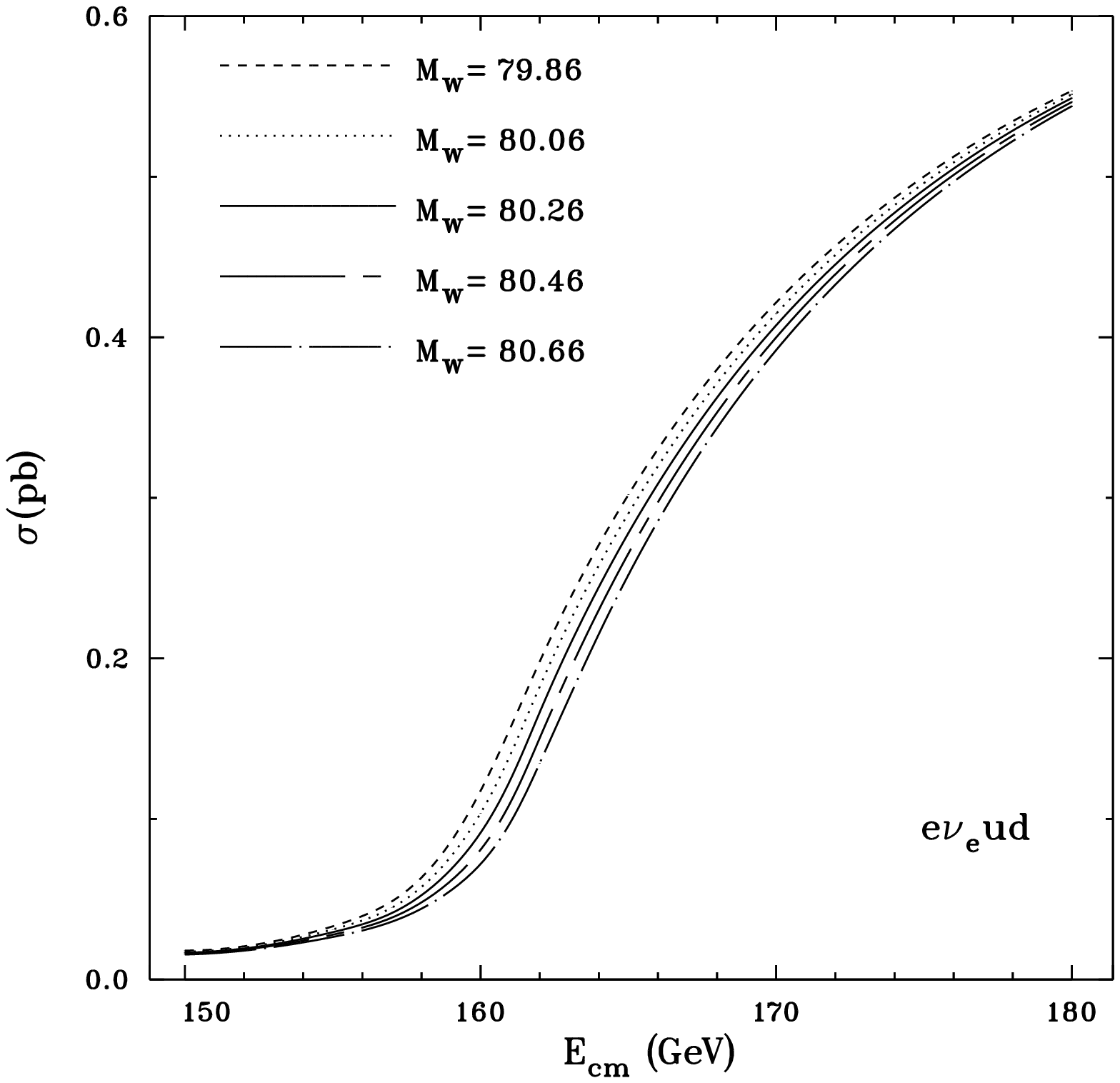,height=18cm,angle=0}
}
Fig.4- Total cross-section for $e^-\bar{\nu}_eu\bar{d}$ process, 
with canonical cuts, versus $\sqrt{s\,}$, from 150 GeV to 205 GeV, 
for different input parameter $\wm$ values.\par

\newpage
\centerline{
\epsfig{figure=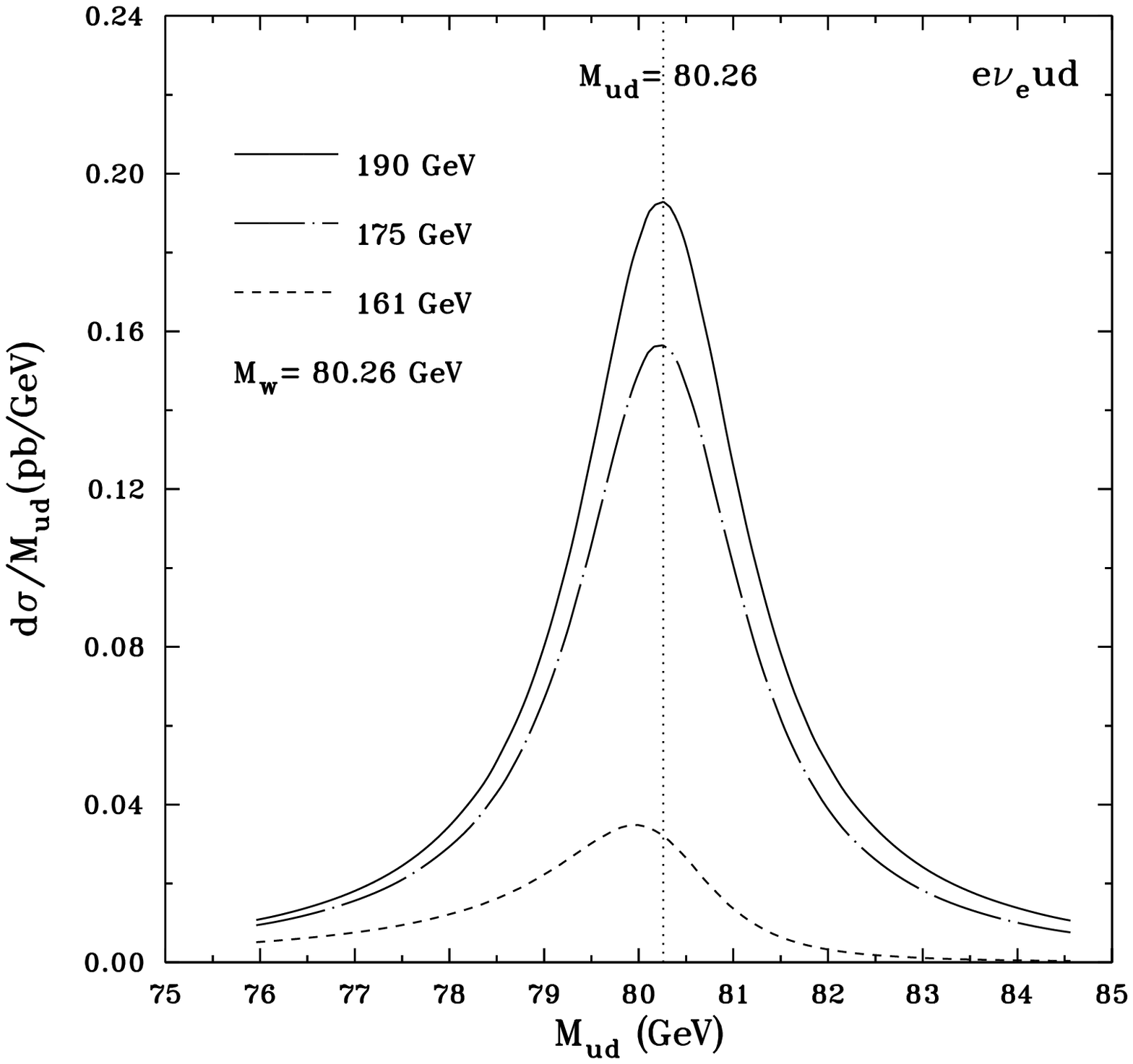,height=18cm,angle=0}
}
Fig.5- Invariant mass distribution of $u\bar{d}$ in the 
$e^-\bar{\nu}_eu\bar{d}$ process, with canonical cuts, at 
$\sqrt{s\,}=$161 GeV (dashed line), 175 GeV (chaindot line)
and 190 GeV (solid line). 

\newpage
\centerline{
\epsfig{figure=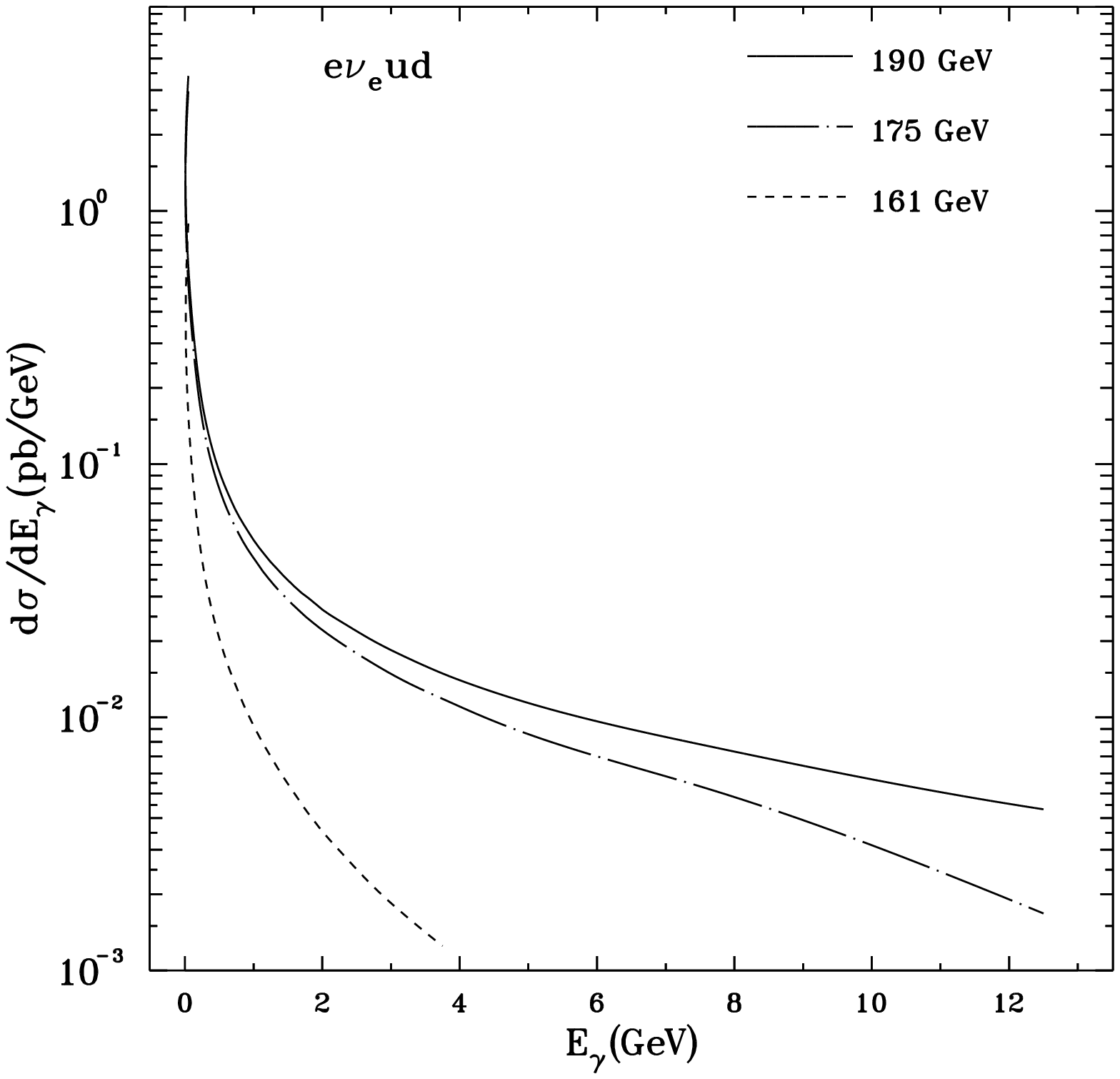,height=18cm,angle=0}
}
Fig.6- Energy spectrum of initial state photons in 
$e^-\bar{\nu}_eu\bar{d}$ 
process, with canonical cuts, at $\sqrt{s\,}=$161 GeV (dashed line), 
175 GeV (chaindot line) 
and 190 GeV (solid line), for $\wm = 80.26\,$ GeV.

\newpage
\centerline{
\epsfig{figure=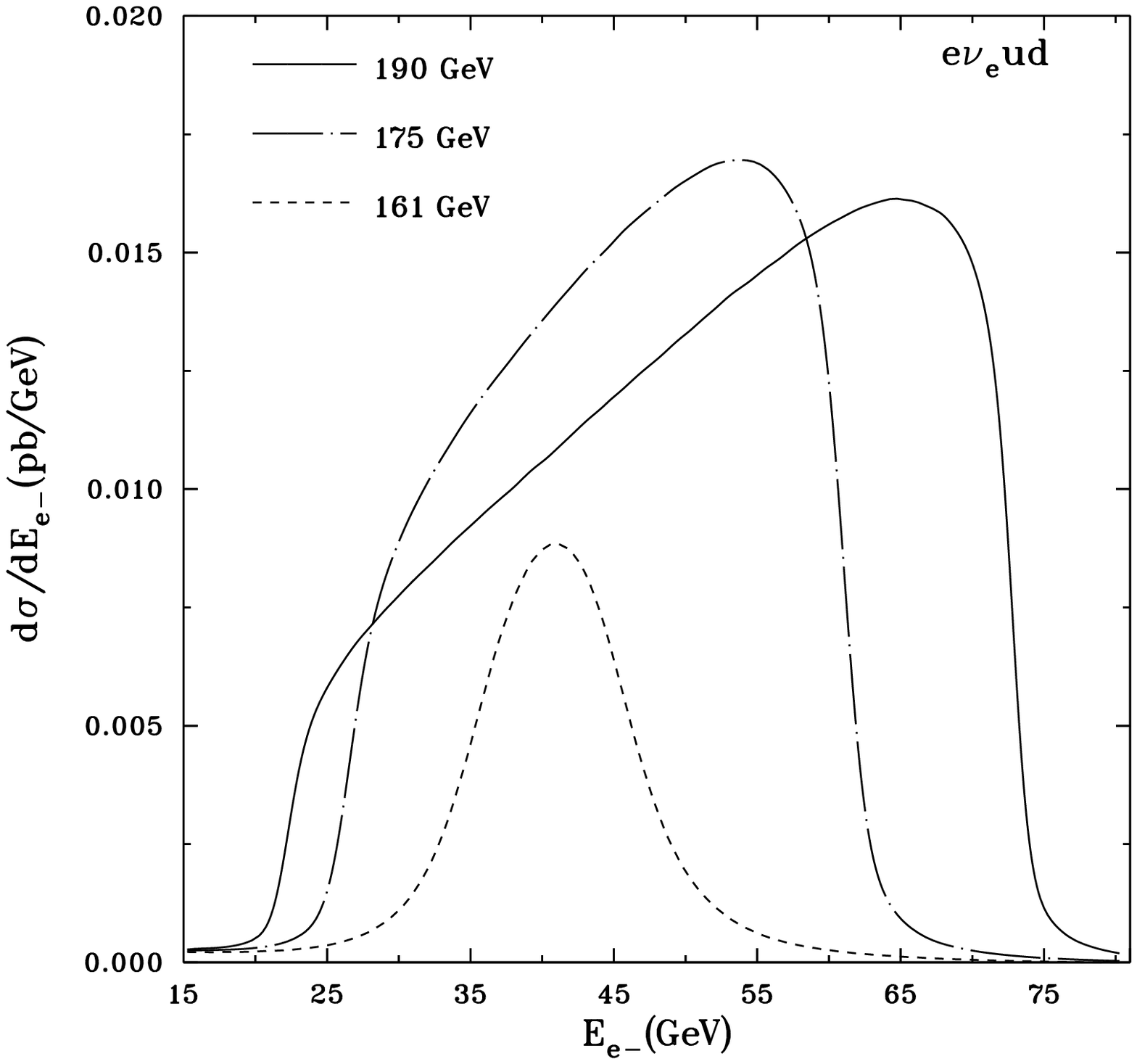,height=18cm,angle=0}
}
Fig.7- Distribution of the final state electron energy in the
$e^-\bar{\nu}_eu\bar{d}$ process, with canonical cuts, at 
$\sqrt{s\,}=$161 GeV (dashed line), 175 GeV
(chaindot line) and 190 GeV (solid line), for $\wm = 80.26\,$ GeV.

\newpage
\centerline{
\epsfig{figure=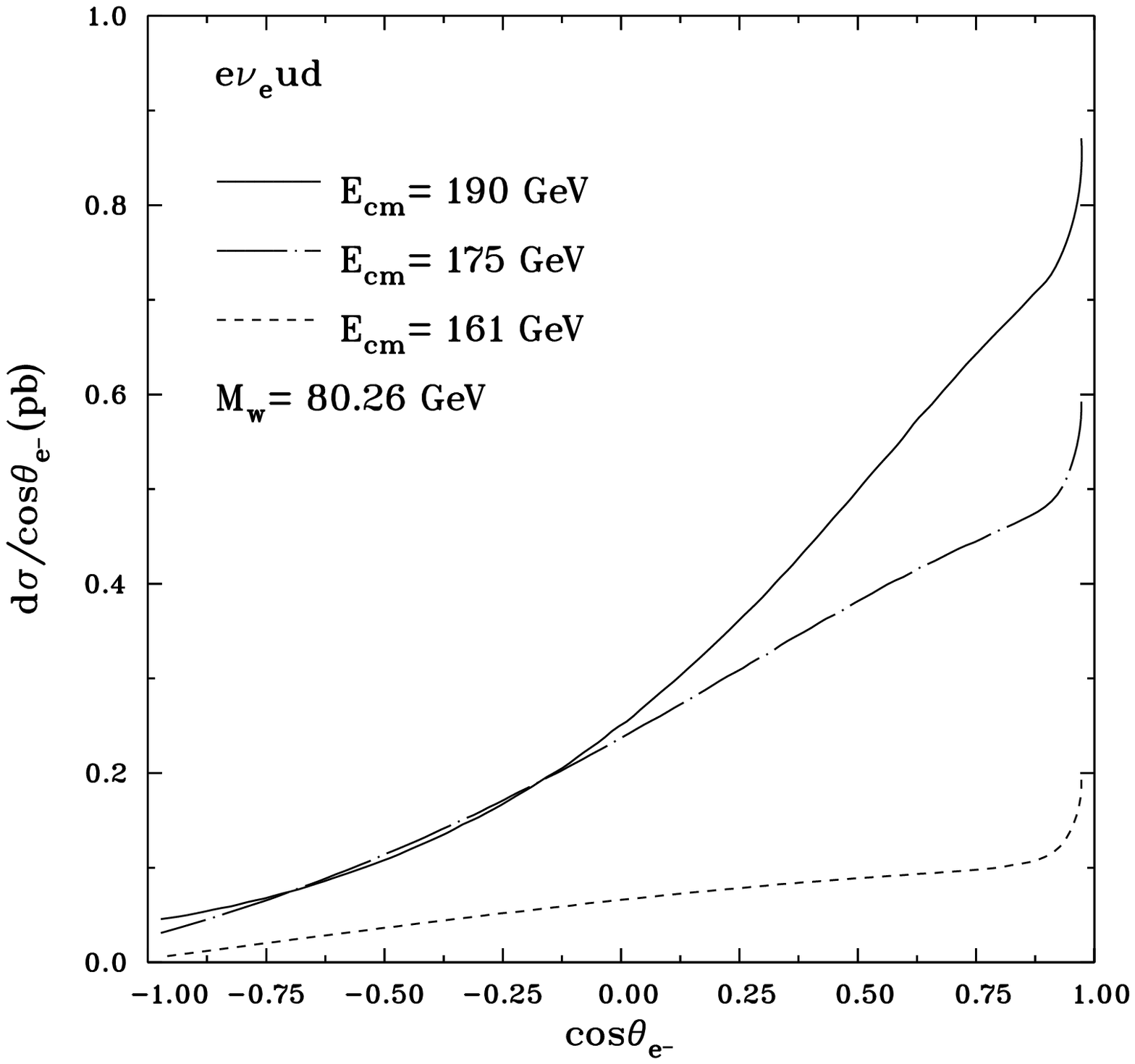,height=18cm,angle=0}
}
Fig.8-  Angular distribution of the final state electron in the
$e^-\bar{\nu}_eu\bar{d}$ process at $\sqrt{s\,}=$161 GeV (dashed line), 175 GeV
(chaindot line) and 190 GeV (solid line). 
Canonical cuts are applied.

\newpage
\centerline{
\epsfig{figure=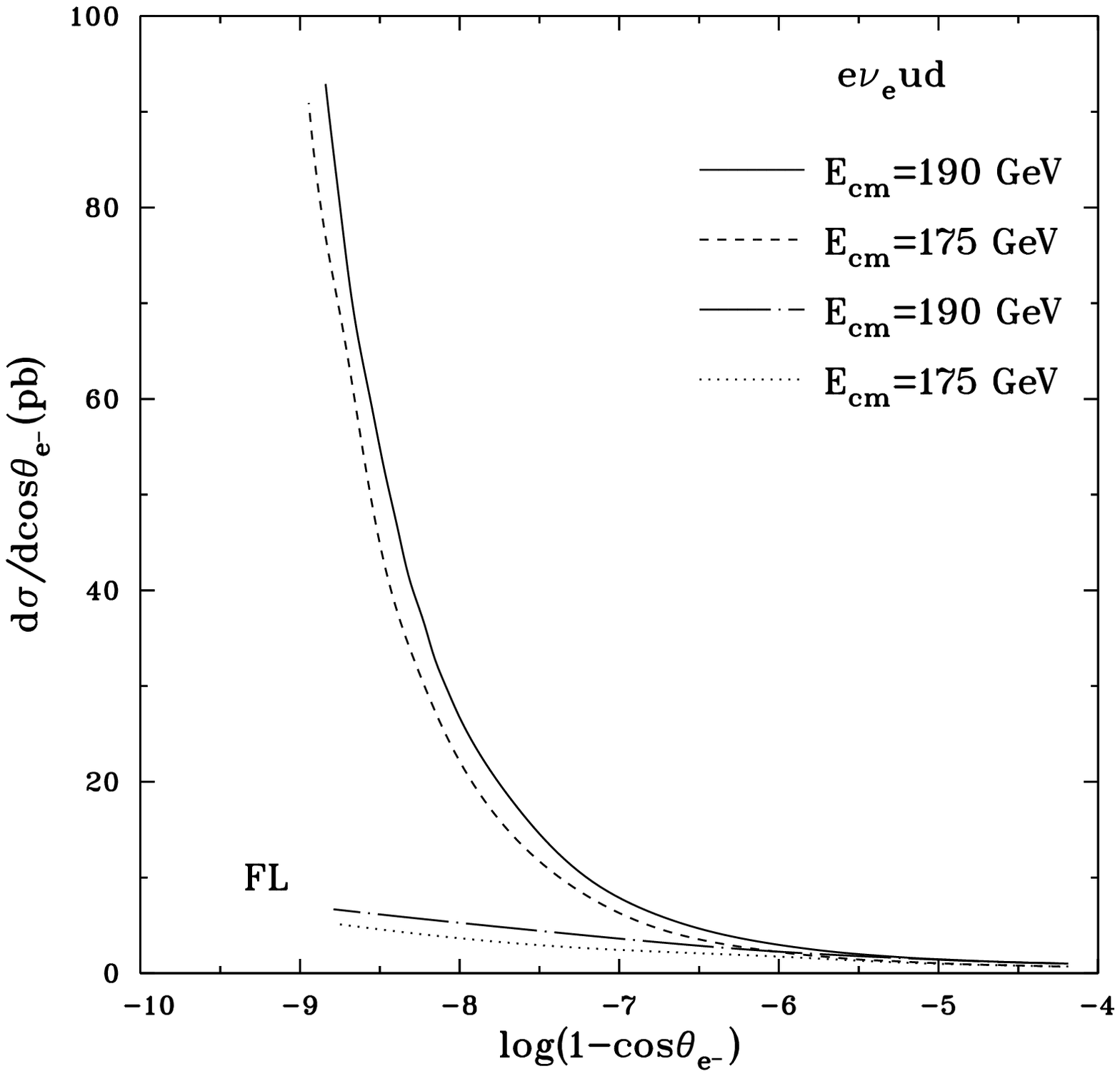,height=18cm,angle=0}
}
Fig.9- Angular distribution of the final state electron in
$e^-\bar{\nu}_eu\bar{d}$, with $\wm=$ 80.26 GeV, 
as a function of 
$log(1-cos\vartheta_{e^-})$, for 10$^o\geq\vartheta_e\geq$1$^o$,
at $\sqrt{s\,}=$175 GeV (dashed line) and 190 GeV (solid line). The dotted
and chaindot curves include the Fermion-Loop scheme at $\sqrt{s\,}=$175 GeV 
and $\sqrt{s\,}=$190 GeV respectively. 
Canonical cuts are applied.

\newpage
\centerline{
\epsfig{figure=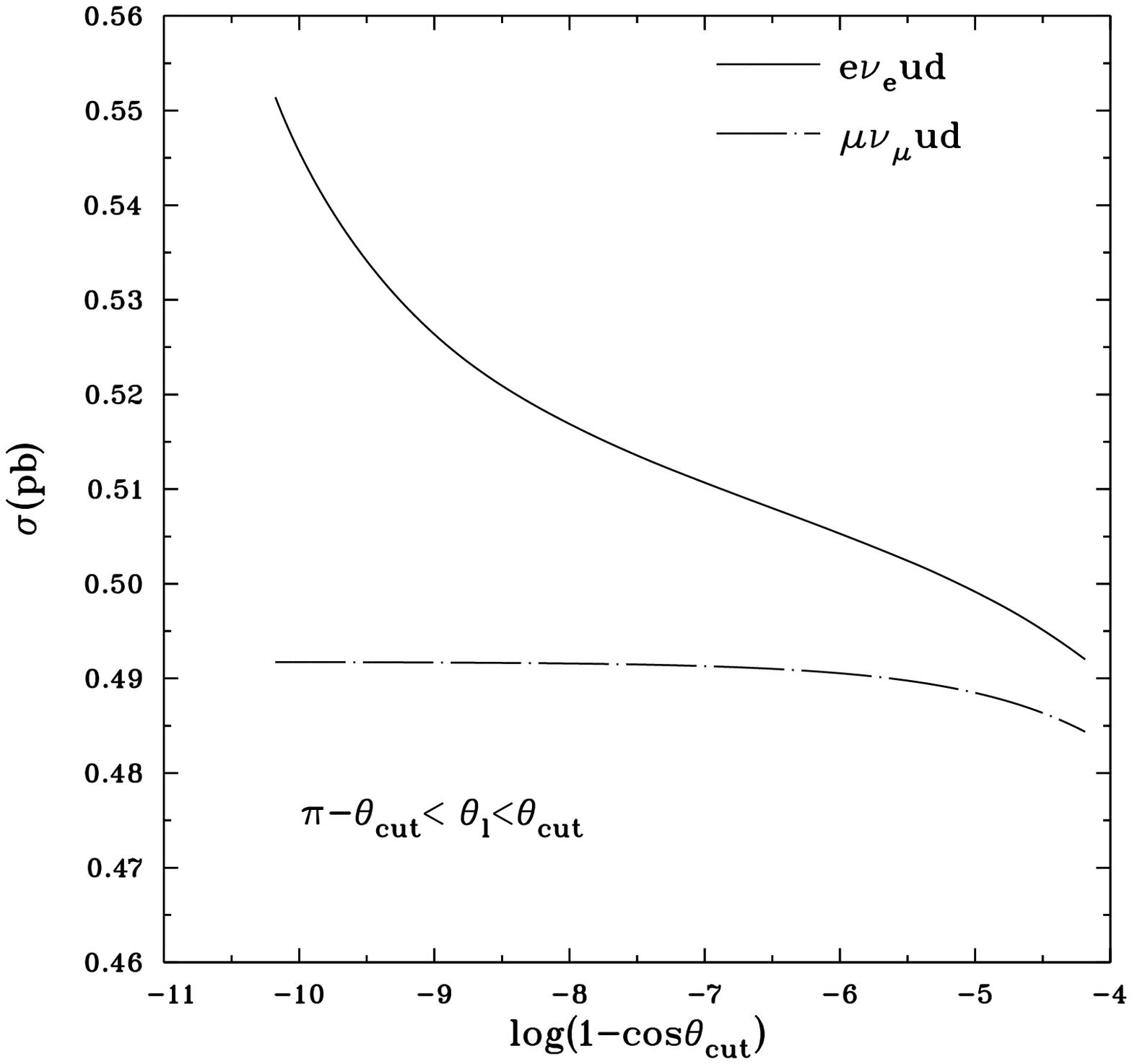,height=18cm,angle=0}
}
Fig.10-  Total cross-section for the semi-leptonic processes
$e^-\bar{\nu}_eu\bar{d}$
 (solid line) and $\mu^-\bar{\nu}_{\mu}u\bar{d}$ (chaindot line) as a function
 of $log(1-cos\vartheta_{cut})$. $\vartheta_{cut}$ the cut imposed on
 the charged lepton scattering angle. The remaining canonical cuts are 
preserved.
$\wm = 80.26\,$ GeV.

\newpage
\centerline{
\epsfig{figure=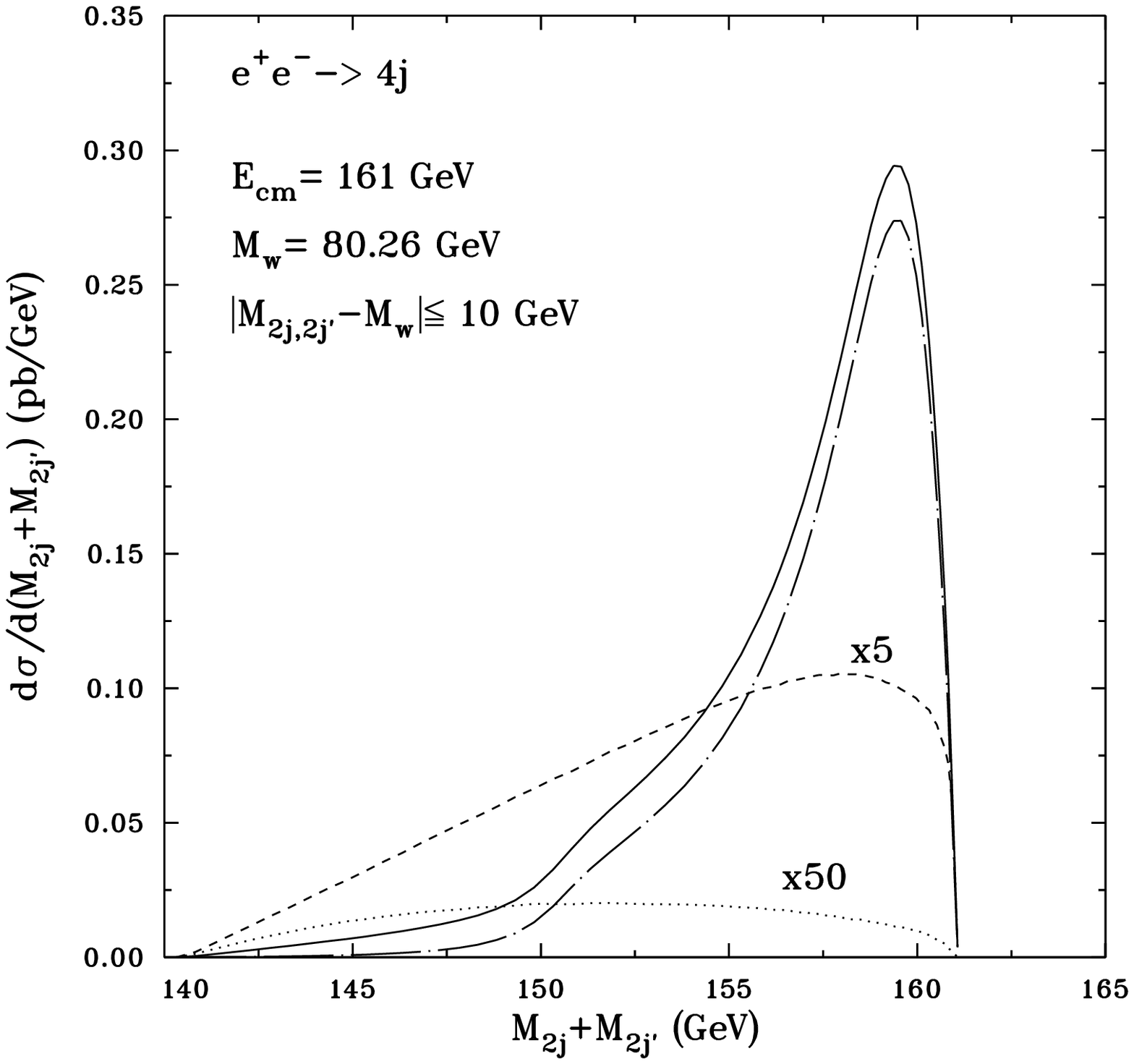,height=18cm,angle=0}
}
Fig.11-  Distribution of the {\it{sum}} of two invariant masses 
in the fully 
hadronic channel at $\sqrt{s\,}=$161 GeV. The chaindot curve corresponds to 
 the two invariant masses from $W^{*\pm}$. The dashed one 
represents the {\it{background}} ( magnified by a factor of 5 ) from 
two non-resonant invariant masses in CC11 and Mix43 processes, counted with 
their
molteplicity. The dotted curve corresponds to the NC {\it{background}} (
magnified by a factor 50 ).
The solid to {\it{signal}}+{\it{total background}}.
For each {\it{sum}},  the two invariant masses lie within 10
GeV  from $\wm$ 

\newpage
\centerline{
\epsfig{figure=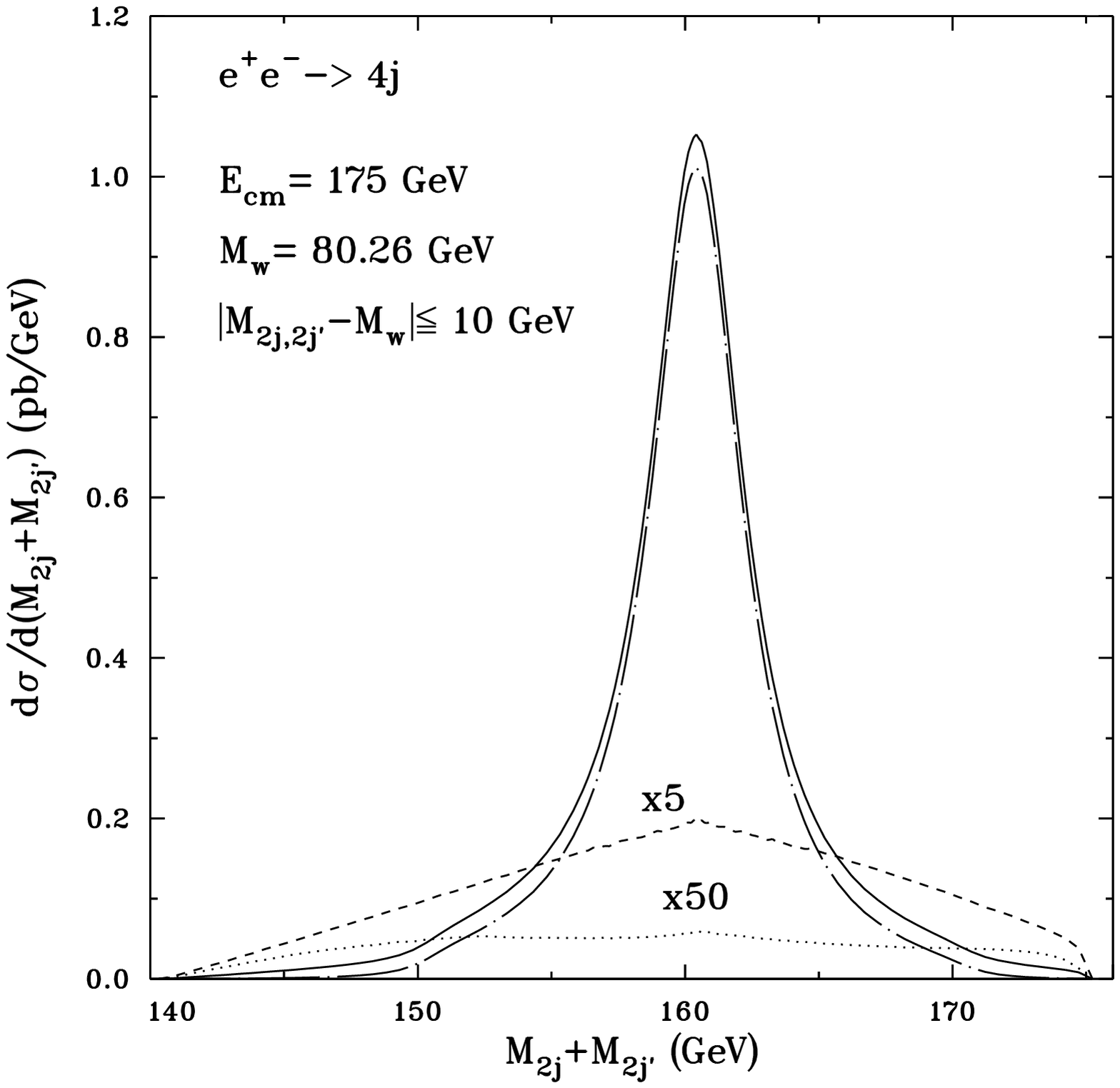,height=18cm,angle=0}
}
Fig.12-  Same distribution as in Fig.11, with $\sqrt{s\,}=$175 GeV.

\newpage
\centerline{
\epsfig{figure=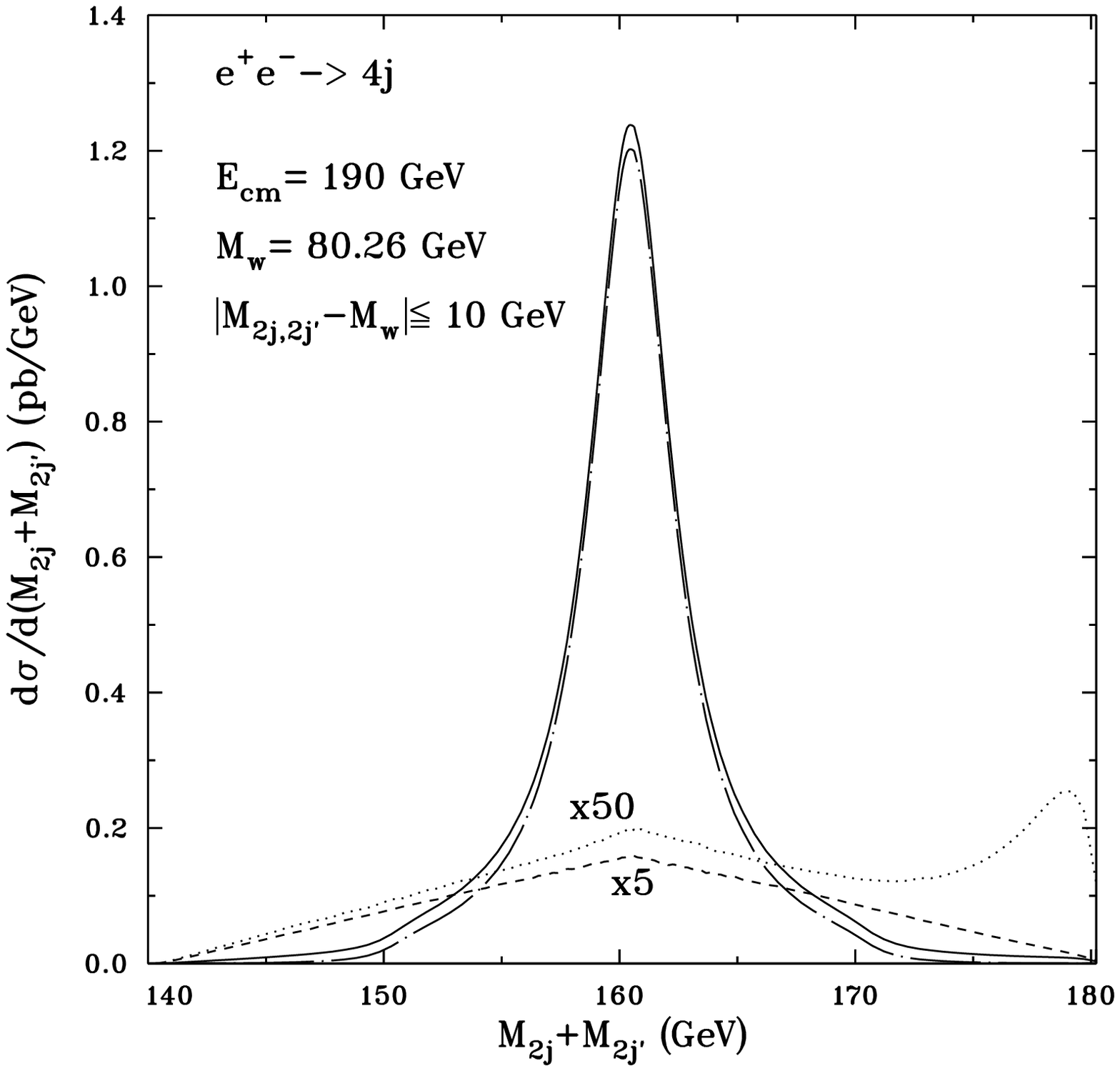,height=18cm,angle=0}
}
Fig.13-  Same distribution as in Fig.11, with $\sqrt{s\,}=$190 GeV.

\end{document}